\documentclass[aps,12pt,eqsecnum,tightenlines,showpacs,nofootinbib,endfloats,amsmath,amssymb]{revtex4}
\usepackage{graphicx}
\usepackage{bm}

\begin{document}

\topmargin -2cm
 \author{A.A. Garcia--Diaz}
 \altaffiliation{aagarcia@fis.cinvestav.mx}
\author{\,G. Gutierrez--Cano }
 \affiliation{Departamento~de~F\'{\i}sica,
 ~Centro~de~Investigaci\'on~y~de~Estudios~Avanzados~del~IPN,\\
 Apdo. Postal 14-740, 07000 M\'exico DF, M\'exico.\\}

 \title{Low energy 2+1 string gravity; black hole solutions}
 \date{\today}

\begin{abstract}
In this report a detailed derivation of the dynamical equations for
an $n$--dimensional heterotic string theory of the Horowitz type is
carried out in the string frame and in the Einstein frame too. In particular, the
dynamical equations of the three dimensional string theory are explicitly given.
The relation of the Horowitz--Welch and  Horne--Horowitz string black
hole solution is exhibited. The Chan--Mann charged dilaton solution is derived and
the subclass of string solutions field is
explicitly identified. The stationary generalization, via $SL(2,R)$
transformations, of the static (2+1) Horne--Horowitz string black
hole solution is given.

\vspace{0.5cm}
\end{abstract}
\maketitle\tableofcontents

\section{$n$--dimensional heterotic string
dynamical equations}\label{StringEqHOR}

Following Horowitz~\cite{Horowitz92}, in this contribution we
reproduce the field equations ``for a part of the low energy
action'' to a $n$--dimensional heterotic string
theory\label{heteroticstring} described by a metric $g_{\mu\nu}$, a scalar field $\Phi$, a
Maxwell field $F_{\mu\nu}$, and a three--form $H_{\mu\nu\lambda}$.
The three--form $\bm H$\label{heteroticH} is related to the
two--form potential $\bm B$ and a gauge field $A_{\mu}$ through $\bm
{H=dB} -a\bm{\,A\wedge dF},$ where $a$ is a constant to be adjusted
at the end for final results. In this text to denote the number of
dimensions is used $n$ instead of $D$. Moreover $\Lambda$ is
reserved for the standard cosmological constant, whereas
$\Lambda_{H}$, and $\Lambda_{CM}=-\Lambda$, denote the
$\Lambda'$s used by Horowitz~\cite{Horowitz92} and Chan and
Mann~\cite{Chan:1995wj} respectively.
 \\

\subsection{String frame}

 The corresponding heterotic
string action\label{actionHor} for dimension $n$,
$S=\int{d^{\,n}x\mathfrak {L}},$  is given
by
\begin{eqnarray}\label{actionHor}
S=\int{d^{\,n}x\sqrt{- g}{\rm e}^{-2\Phi}\left[{ R} -2
\Lambda+U(\Phi)+4 (\nabla\Phi)^2 -F^{2} -\frac{1}{12}H^{2}\right]}.
\end{eqnarray}
Variations with respect to the
metric\label{Variations with} give:
\begin{subequations}\label{actionHorEINS}
\begin{eqnarray}
\frac{\delta \sqrt{- g}}{\delta { g}^{\mu\nu}} =-\frac{1}{2}\sqrt{-
g}{ g}_{\mu\nu},
\end{eqnarray}
\begin{eqnarray}\label{VariationRHor1}
\frac{\delta \sqrt{- g}{ R}} {\delta { g}^{\mu\nu}}=\frac{\delta
\sqrt{- g}{ g }^{\alpha\beta} { R}_{\alpha\beta}}{\delta {
g}^{\mu\nu}}&& =\sqrt{- g}({ R}_{\mu\nu}-\frac{1}{2}{ g}_{\mu\nu}{
R}) +\sqrt{- g}{ g }^{\alpha\beta} \frac{\delta {
R}_{\alpha\beta}}{\delta { g}^{\mu\nu}}\nonumber\\&& =\sqrt{- g}\,{
G}_{\mu\nu}+\sqrt{- g}{ g }^{\alpha\beta}\frac{\delta {
R}_{\alpha\beta}}{\delta { g}^{\mu\nu}},
\end{eqnarray}
\begin{eqnarray}\label{VariationRHor2}
\frac{\delta }{\delta { g}^{\mu\nu}} \sqrt{-
g}(\nabla\Phi)^2=\frac{\delta } {\delta { g}^{\mu\nu}}\sqrt{- g}{
g}^{\alpha\beta} \nabla_{\alpha}\,\Phi\,\nabla_{\beta}\,\Phi
=\sqrt{- g}(\nabla_{\mu}\Phi\,\nabla_{\nu}\Phi -\frac{1}{2}{
g}_{\mu\nu} (\nabla\Phi)^2),
\end{eqnarray}
\begin{eqnarray}\label{VariationRHor3}
\frac{\delta }{\delta { g}^{\mu\nu}} \sqrt{- g}\,F^2=\frac{\delta
}{\delta { g}^{\mu\nu}} \sqrt{- g}{{ g}^{\alpha\lambda}{
g}^{\beta\rho} F_{\alpha\beta}F_{\lambda\rho}}=2\, \sqrt{- g}({
g}^{\beta\rho} F_{\mu\beta}F_{\nu\rho} -\frac{1}{4}{
g}_{\mu\nu}F^2),
\end{eqnarray}
\begin{eqnarray}\label{VariationRHor4}
\frac{\delta}{\delta { g}^{\mu\nu}} \sqrt{- g}H^{2} =\frac{\delta
}{\delta { g}^{\mu\nu}} \left(\sqrt{- g}{{ g}^{\alpha\sigma}{
g}^{\beta\rho} { g}^{\gamma\lambda}H_{\alpha\beta\gamma}
H_{\sigma\rho\lambda}}\right)=3\,\sqrt{- g}
(H_{\mu\alpha\beta}{H_{\nu}}^{\alpha\beta} -\frac{1}{6}{
g}_{\mu\nu}\,H^{2}).
\end{eqnarray}
\end{subequations}
As far as to the term $\sqrt{- g}{ g
}^{\alpha\beta} {\delta { R}_{\alpha\beta}}$ in
Eq.~(\ref{VariationRHor1}) is concerned, it is easy to establish its
proportionality to a divergence, ${\it v^\mu}_{;\mu}$: in standard
textbooks one find the variation of the Riemann tensor in terms of the variation of the
Christoffel symbols
\begin{eqnarray}
\delta{R^\lambda}_{\mu\kappa\nu}=(\delta{\Gamma^{\lambda}}_{\mu\nu})_{;\kappa}-(\delta
{\Gamma^{\lambda}}_{\mu\kappa})_{;\nu}
\end{eqnarray}
then, since ${ R}_{\mu\nu}= { R^\kappa}_{\mu\kappa\nu}$, one has
\begin{eqnarray}
&& { g}^{\mu\nu}\delta{{ R}_{\mu\nu}}=({ g}^{\mu\lambda}\delta
{{\Gamma}^{\nu}}_{\mu\lambda})_{;\nu}-({ g}^{\mu\nu} \delta{{
\Gamma}^{\lambda}}_{\mu\lambda})_{;\nu} =({ g}^{\mu\lambda}\delta
{{\Gamma}^{\nu}}_{\mu\lambda} -{
g}^{\mu\nu}\delta{{\Gamma}^{\lambda}}_{\mu\lambda} )_{;\nu}=:{\it
v^\nu}_{;\nu}=
\nonumber\\
&& =\frac{1}{\sqrt{-{ g}}} \left[\sqrt{-{ g}}\left({
g}^{\mu\lambda}\delta {{\Gamma}^{\nu}}_{\mu\lambda}-{ g}^{\mu
\nu}\delta {{\Gamma}^{\lambda}}_{\mu\lambda}\right)\right]_{,\nu}.
\end{eqnarray}
This term, taking into account the
multiplicative factor $\exp{(-2\Phi)}$, will give rise in the
Einstein's equation to second order derivatives of $\Phi$ due first
to the covariant derivative acting on $\delta{\Gamma}$ when forming
a new divergence, and second to the covariant derivative acting on
variations of the metric $\delta{ g}$ for $\delta{\Gamma}$ expressed in
terms of $\delta{ g}$ according to~\footnote{Acting with  the
variation operation $\delta$ on the metric tensor and its
derivatives in the expression ${{\Gamma}^\mu}_{\alpha\beta},$ taking
into account the commutativity of  $\delta$ and partial derivatives
${\partial_{,i}},$ and ${ g}^{\mu\nu} {\delta g}_{\sigma\nu}=-{
\delta g}^{\mu\nu}{ g}_{\sigma\nu},$ one has
\begin{eqnarray*}
&&{{\Gamma}^\mu}_{\alpha\beta} =\frac{1}{2}{ g}^{\mu\nu}\left[{
g}_{\nu\beta,\alpha} +{ g}_{\alpha\nu,\beta}-{
g}_{\alpha\beta,\nu}\right] =\frac{1}{2}{
g}^{\mu\nu}\left[\alpha\beta,\nu\right],
\left[\alpha\beta,\nu\right]=2{ g}_{\mu\nu}{{\Gamma}^\mu}_{\alpha\beta},\nonumber\\
&&
 \delta{{\Gamma}^\mu}_{\alpha\beta}=\frac{1}{2}{\delta g}^{\mu\nu}
 \left[\alpha\beta,\nu\right]
 +\frac{1}{2}{ g}^{\mu\nu}\delta\left[\alpha\beta,\nu\right]
 ={\delta g}^{\mu\nu}{ g}_{\sigma\nu}{{\Gamma}^\sigma}_{\alpha\beta}
 +\frac{1}{2}{ g}^{\mu\nu}\delta\left[\alpha\beta,\nu\right]\nonumber\\
 &&
 =-{ g}^{\mu\nu}{\delta g}_{\sigma\nu}
 {{\Gamma}^\sigma}_{\alpha\beta}
 +\frac{1}{2}{ g}^{\mu\nu}\left[(\delta{ g}_{\nu\beta})_{,\alpha}
 +(\delta{ g}_{\alpha\nu})_{,\beta}-(\delta{ g}_{\alpha\beta})_{,\nu}\right]
 =\delta{{\Gamma}^\mu}_{\alpha\beta}(\ref{VarGammaVarmet}).
\end{eqnarray*}},
\begin{eqnarray}\label{VarGammaVarmet}
 \delta{{\Gamma}^\mu}_{\alpha\beta}
 =\frac{1}{2}{ g}^{\mu\nu}\left(\nabla_{\alpha}\delta{ g}_{\nu\beta}
 +\nabla_{\beta}\delta{ g}_{\alpha\nu}
 -\nabla_{\nu}\delta{ g}_{\alpha\beta}\right),
\end{eqnarray}
which implies
\begin{eqnarray}
 { \it  v^\nu}={ g}^{\nu\mu}{ g}^{\alpha\beta}
 \left(\nabla_{\alpha}\delta{ g}_{\mu\beta}
 -\nabla_{\mu}\delta{ g}_{\alpha\beta}\right)
 =[\delta{ g}_{\alpha\beta}
 ({ g}^{\nu\alpha}{ g}^{\mu\beta}
 -{ g}^{\nu\mu}{ g}^{\alpha\beta})]_{;\mu},
 \end{eqnarray}
therefore,
\begin{eqnarray}
&& e^{-2\Phi}\sqrt{-{ g}}{ g}^{\mu\nu} \delta{{
R}_{\mu\nu}}=e^{-2\Phi} \sqrt{-{ g}}{\it  v^\nu}_{;\nu}
\nonumber\\&&=e^{-2\Phi} \left(\sqrt{-{ g}}{\it
v^\nu}\right)_{,\nu}=\left(e^{-2\Phi}\sqrt{-{ g}}{\it
v^\nu}\right)_{,\nu}+2\,\Phi_{,\nu}e^{-2\Phi}\,\sqrt{-{ g}}{\it
v^\nu},
\end{eqnarray}
consequently
\begin{eqnarray}\label{HORphicontribution}
2\Phi_{,\nu}&&e^{-2\Phi}\, \sqrt{-{ g}}\,{\it   v^\nu} =2\sqrt{-{
g}}\left[e^{-2\Phi}\Phi_{,\nu}\delta{ g}_{\alpha\beta} \left({
g}^{\nu\alpha}{ g}^{\mu\beta} -{ g}^{\nu\mu}{
g}^{\alpha\beta}\right)\right]_{;\mu}
\nonumber\\&&-2\left(e^{-2\Phi}\Phi_{,\nu} \right)_{;\mu}\sqrt{-{
g}}{ g}^{\nu\alpha} { g}^{\mu\beta}\delta{ g}_{\alpha\beta}
+2\left(e^{-2\Phi}\Phi_{,\nu} \right)_{;\mu}\sqrt{-{ g}}{
g}^{\nu\mu} { g}^{\alpha\beta}\delta{ g}_{\alpha\beta}\nonumber\\&&
=2\,\sqrt{-{ g}}{\rm {Div}} + 2\sqrt{-{
g}}\left[\left(e^{-2\Phi}\Phi_{,\nu} \right)_{;\mu}-{ g}_{\mu\nu}{
g}^{\alpha\beta}\left(e^{-2\Phi}\Phi_{,\alpha}
\right)_{;\beta}\right]\delta{ g}^{\mu\nu},
\end{eqnarray}
where it has been used $\delta g_{\mu\nu}
=-g_{\mu\alpha}g_{\nu\beta}\delta g^{\alpha\beta}.$ Gathering all
the divergence terms, by applying the Stokes theorem\label{Stokes
theorem}, the integral of the total derivative terms becomes an
integral over the boundary; imposing conditions on the variations
$\delta{ g}_{\alpha\nu}$ and on the covariant derivatives of these
variations, $\nabla_{\mu}\delta{ g}_{\alpha\nu}$, one can avoid the
contributions of the boundary terms. On this respect see
Wald~\cite{Wald:1984book}, Appendix E, and also
Ort{\'i}n~\cite{Ortin:2004book}, Part I, Chapter 4. The second term
in the last line of Eq.(\ref{HORphicontribution}) contributes to
Einstein equations. Therefore $\frac{\delta \mathfrak {L} } {\delta
{ g}^{\mu\nu}}=0$ gives rise to Einstein equations in the form
\begin{eqnarray}\label{HOREins1}
&&{ G}_{\mu\nu}+\frac{1}{2}\, \left(2\Lambda-U(\Phi)\right){
g}_{\mu\nu}+ 2\,\Phi_{,\nu;\mu}-2{ g}_{\mu\nu}{ g}^{\alpha\beta}
\Phi_{,\alpha;\beta}+ 2\,{ g}_{\mu\nu} {
g}^{\alpha\beta}\,\Phi_{,\alpha}\,\Phi_{,\beta} \nonumber\\&&
-2\,\left({ g}^{\beta\rho}F_{\mu\beta}F_{\nu\rho} -\frac{1}{4}{
g}_{\mu\nu}{ F}^2\right)-\frac{1}{4}
\left(H_{\mu\alpha\beta}{H_{\nu}}^{\alpha\beta} -\frac{1}{6}{
g}_{\mu\nu}\,H^{2}\right)=0.
\end{eqnarray}
The variation with respect to the scalar
field $\Phi$, $\frac{\delta \mathfrak {L} }{\delta {\Phi}}
=(\frac{\partial}{\partial \Phi} -\frac{\partial}{\partial x^\alpha}
\frac{\partial}{\partial \Phi_{,\alpha}})\mathfrak {L} ,$
considering that
\begin{eqnarray*}
&&\frac{\partial}{\partial x^\alpha}\left({\rm e}^{-2\Phi}\sqrt{- g}
\frac{\partial}{\partial \Phi_{,\alpha}}({ g}^{\mu\nu}
\Phi_{,\mu}\Phi_{,\nu})\right)={2} \frac{\partial}{\partial
x^\alpha} \left({\rm e}^{-2\Phi}\sqrt{- g}\,{ g}^{\alpha\nu}
\Phi_{,\nu}\right) \nonumber\\&& =2{\rm e}^{-2\Phi}\left(-2\sqrt{-
g}\, \Phi_{,\alpha}\,{ g}^{\alpha\nu}\Phi_{,\nu}
+\frac{\partial}{\partial x^\alpha} (\sqrt{- g}\,{
g}^{\alpha\nu}\Phi_{,\nu})\right) \nonumber\\&& =-4{\rm
e}^{-2\Phi}\sqrt{- g}\, \Phi_{,\alpha}\,{ g}^{\alpha\nu}
\Phi_{,\nu}+2{\rm e}^{-2\Phi} \sqrt{- g}({
g}^{\alpha\nu}\Phi_{;\nu})_{;\alpha},
\end{eqnarray*}
yields the dynamical equation
\begin{eqnarray}\label{HOREins2}
4{ g}^{\alpha\nu}\Phi_{;\nu;\alpha} -4 {
g}^{\mu\nu}\Phi_{,\mu}\Phi_{,\nu}+{ R} -2\Lambda-{
F}^{2}-\frac{1}{12}H^2 +U(\Phi)-\frac{1}{2}\frac{d\,U}{d\Phi}=0.
\end{eqnarray}
The field equation for $B_{\mu\nu}$ arises
from the variation of $H^2$, where
$$H_{\mu\nu\lambda}=3\,B_{[\mu\nu,\lambda]}
-a\,A_{[\mu}\,F_{\nu\lambda]} \rightarrow\frac{\partial
H_{\alpha\beta\gamma}} {B_{\mu\nu,\sigma}}=3\,\delta^\mu_{[\alpha}
\delta^\nu_{\beta}\delta^\sigma_{\gamma]},$$
consequently
\begin{eqnarray}\label{HOREins3}
\frac{\partial }{\partial B_{\mu\nu,\sigma}}H^2 =\frac{\partial
H_{\alpha\beta\gamma}} {\partial B_{\mu\nu,\sigma}}
\frac{\partial}{\partial H_{\alpha\beta\gamma}}{{ g}^{\tau\sigma} {
g}^{\kappa\rho}{ g}^{\lambda\xi}H_{\tau\kappa\xi}
H_{\sigma\rho\lambda}} =2\frac{\partial
H_{\alpha\beta\gamma}}{B_{\mu\nu,\sigma}}
H^{\alpha\beta\gamma}=3!H^{\mu\nu\sigma}
\end{eqnarray}
therefore, for $\frac{\delta }{\delta {
B_{\mu\nu}}}\mathfrak {L} =(\frac{\partial }{\partial { B_{\mu\nu}}}
-\frac{\partial}{\partial{x^{\sigma}}} \frac{\partial }{\partial {
B_{\mu\nu,\sigma}}}) \mathfrak {L},$ one gets
\begin{eqnarray}\label{DynEDqforB}
\frac{\delta }{\delta { B_{\mu\nu}}} \sqrt{- g}{\rm
e}^{-2\Phi}H^2=-3!\frac{\partial}{\partial x^\sigma } \left(\sqrt{-
g}{\rm e}^{-2\Phi}H^{\mu\nu\sigma}\right)
\rightarrow\nabla_{\sigma}({\rm e}^{-2\Phi} H^{\mu\nu\sigma})=0.
\end{eqnarray}
Finally, the electromagnetic Maxwell
equations are derived for $A_{\alpha}$ related with the
electromagnetic field through $F_{\mu\nu}
=F_{\mu\nu}(A_{\alpha})=A_{\nu,\mu}-A_{\mu,\nu}
=2\partial_{[\mu}A_{\nu]},$
\begin{eqnarray}\label{HOREinsMax}
\frac{\delta}{\delta A_{\epsilon}}\mathfrak {L}&&
=-(\frac{\partial}{\partial A_{\epsilon}} -\frac{\partial}{\partial
x^\sigma} \frac{\partial}{\partial A_{\epsilon,\sigma}}) [\sqrt{-
g}{\rm e}^{-2\Phi}(F^2+\frac{1}{12}H^2)] =-\frac{1}{12}\,\sqrt{-
g}{\rm e}^{-2\Phi} \frac{\partial H_{\alpha\beta\gamma}} {\partial
A_{\epsilon}}\frac{\partial}{\partial
H_{\alpha\beta\gamma}}H^2\nonumber\\
&& +\frac{\partial}{\partial x^\sigma} (\sqrt{- g}{\rm e}^{-2\Phi}
\frac{\partial F_{\alpha\beta} }{\partial A_{\epsilon,\sigma}}
\frac{\partial}{\partial F_{\alpha\beta}} F^2)+\frac{1}{12}
\frac{\partial}{\partial x^\sigma}(\sqrt{- g}{\rm e}^{-2\Phi}
\frac{\partial H_{\alpha\beta\gamma}}{\partial A_{\epsilon,\sigma}}
\frac{\partial}{\partial H_{\alpha\beta\gamma}}H^2)
\end{eqnarray}
taking into account that
\begin{eqnarray}\label{HOREinsMax2}
\frac{\partial}{\partial F_{\alpha\beta}} F^2
=2F^{\alpha\beta},\,
\frac{\partial F_{\alpha\beta}}{\partial
A_{\epsilon,\lambda}}
=-2\delta^{\epsilon}_{[\alpha}
\delta^{\lambda}_{\beta]},\,\frac{\partial}{\partial
H_{\alpha\beta\gamma}}H^2
=2H^{\alpha\beta\gamma}\nonumber\\
\end{eqnarray}
together with
$$\frac{\partial}{\partial A_{\epsilon}}H_{\mu\nu\lambda}
=-a\delta^\epsilon_{[\mu}\,F_{\nu\lambda]}$$
$$\frac{\partial}{\partial A_{\epsilon,\sigma}}H_{\alpha\beta\gamma}
=-a\,\frac{\partial}{\partial
A_{\epsilon,\sigma}}A_{[\alpha}\,F_{\beta\gamma]}=
2\,a\,(A_{[\alpha}\,\delta^\epsilon_{\beta}\delta^\sigma_{\gamma]})$$
one has for $\frac{\delta \mathfrak {L}
}{\delta A_{\epsilon}}$ the expression
\begin{eqnarray}\label{HOREinsMax3}
\frac{\delta \mathfrak {L} }{\delta A_{\epsilon}}
&&=-4\frac{\partial}{\partial x^\lambda} (\sqrt{- g}\,{\rm
e}^{-2\Phi}{ F}^{\epsilon\lambda}) +a\frac{1}{6}\,\sqrt{- g}{\rm
e}^{-2\Phi}F_{\nu\lambda}\, { H}^{\epsilon\nu\lambda}
 \nonumber\\&&
+a\frac{1}{3}\frac{\partial}{\partial x^\sigma}
\left(A_{[\alpha}\,\delta^\epsilon_{\beta}\delta^\sigma_{\gamma]}
\sqrt{- g}{\rm e}^{-2\Phi}{ H}^{\alpha\beta\gamma}\right)
\nonumber\\&&=4\,\sqrt{- g}\left({\rm e}^{-2\Phi}{
F}^{\lambda\epsilon}\right)_{;\lambda} +a\frac{1}{3}\sqrt{- g}{\rm
e}^{-2\Phi}F_{\nu\lambda}\,{ H}^{\epsilon\nu\lambda}=0,
\end{eqnarray}
where the equation (\ref{DynEDqforB}) has been used.\\

Summarizing, the field dynamical
equations\label{field dynam} are
\begin{subequations}\label{einsteinHORdyn}
\begin{eqnarray}\label{einsteinHOR1}
 {G}_{\mu\nu}&&-\frac{1}{2}{ g}_{\mu\nu}\,
(-2\Lambda+U)+2\nabla_{\mu}\nabla_{\nu}\Phi -2{
g}_{\mu\nu}{\nabla}^{2}\Phi +2{ g}_{\mu\nu}
(\nabla\,\Phi)^2\nonumber\\&& -2\,(F_{\mu\alpha}F_{\nu\beta}{
g}^{\alpha\beta} -\frac{1}{4}{ g}_{\mu\nu}{
F}^2) -\frac{1}{4}(H_{\mu\alpha\beta}{H_{\nu}}^{\alpha\beta}
-\frac{1}{6}{g}_{\mu\nu}{ H}^2)=0,
\end{eqnarray}
\begin{eqnarray}\label{scalarHORscaS}
4\nabla^{2}\,\Phi
-4(\nabla\,\Phi)^2+R-2\,\Lambda-{
F}^{2}-\frac{1}{6}H^2 +U(\Phi)-\frac{1}{2}\frac{d\,U}{d\,\Phi}=0,
\end{eqnarray}
\begin{eqnarray}\label{BfieldHOR}
\nabla_{\sigma}({\rm e}^{-2\Phi}H^{\mu\nu\sigma})=0,
\end{eqnarray}
\begin{eqnarray}\label{AfieldHOR}
\nabla_{\lambda}\left({\rm e}^{-2\Phi}{ F}^{\lambda\epsilon}\right)
+a\frac{1}{12}{\rm e}^{-2\Phi}F_{\nu\lambda}\,{
H}^{\epsilon\nu\lambda}=0.
\end{eqnarray}
\end{subequations}
The last equation differs in sign from the
corresponding Horowitz's equation (2.10b).\\
By constructing (\ref{einsteinHOR1}) one evaluates ${ R}$,
\begin{eqnarray}\label{einsteinHORscal}
&&R=\frac{2}{n-2}\left(n \Lambda-\frac{n}{2}U-2(n-1)\nabla^2\Phi +2n({\nabla}\Phi)^2 +\frac{n-4}{2}{
F}^2+\frac{n-6}{24}{ H}^2\right),
\end{eqnarray}
which replaced again into
(\ref{einsteinHOR1}) and (\ref{scalarHORscaS}) allows one to rewrite the set of dynamical
equations as
\begin{subequations}\label{FieldEQstring}
\begin{eqnarray}\label{einsteinHORsumm}
 &&{R}_{\mu\nu}=-2\nabla_{\mu}\nabla_{\nu}\Phi+2F_{\mu\alpha}F_{\nu\beta}{
g}^{\alpha\beta
}+\frac{1}{4}\,H_{\mu\alpha\beta}{H_{\nu}}^{\alpha\beta}
\nonumber\\&&+
\frac{2}{n-2}{ g}_{\mu\nu}\,\left(\Lambda-\frac{1}{2}U-\nabla^2\Phi+2({\nabla}\Phi)^2
 -\frac{1}{2}F^2-\frac{1}{12}{ H}^2\right),
\end{eqnarray}
\begin{eqnarray}\label{tildeEinsHorScal}
{2}\nabla^2\Phi -4 (\nabla\Phi)^2 -2\Lambda+\,{
F}^{2}+\frac{1}{6}H^2+U
+\frac{n-2}{4}\frac{d\,U}{d\,\Phi} =0,
\end{eqnarray}
\begin{eqnarray}\label{BfieldHOR}
\nabla_{\sigma}({\rm e}^{-2\Phi}H^{\mu\nu\sigma})=0,
\end{eqnarray}
\begin{eqnarray}\label{AfieldHOR}
\nabla_{\lambda}\left({\rm e}^{-2\Phi}{ F}^{\lambda\epsilon}\right)
+a\frac{1}{12}{\rm e}^{-2\Phi}F_{\nu\lambda} \,{
H}^{\epsilon\nu\lambda}=0.
\end{eqnarray}
\end{subequations}

\subsection{Einstein frame}

On the other hand, to pass to the Einstein frame
description\label{Einsteinframe} of this low energy string theory,
one accomplishes a conformal transformation\label{conformaltran} of
the form
\begin{eqnarray}
{\tilde g}_{\mu\nu}={\rm e}^{2\sigma}{g}_{\mu\nu}, {\tilde
g}^{\mu\nu}={\rm e}^{-2\sigma}{g}^{\mu\nu}\rightarrow {\tilde
g}={\rm e}^{2\,n\sigma}{g}, \rm dim=n,
\end{eqnarray}
which transforms the action $S$ (\ref{actionHor}), considered as the
barred one, taking into account that ${\tilde F}^2={\tilde
g}^{\mu\alpha}{\tilde g}^{\nu\beta} F_{\mu\nu}F_{\alpha\beta}={\rm
e}^{-4\sigma}{ g}^{\mu\alpha}{ g}^{\nu\beta}
F_{\mu\nu}F_{\alpha\beta}={\rm e}^{-4\sigma}F^2,$ and ${\tilde
H}^2={\rm e}^{-6\sigma}H^2,$ to the form
\begin{eqnarray}\label{actioMiddle}
S=\int{d^{\,n}x\sqrt{-g}{\rm e}^{(n\,\sigma-2\,\sigma
-2\Phi)}\left[{\rm e}^{2\sigma}{\tilde R}+{\rm e}^{2\,
\sigma}(-2\Lambda+U)+4(\nabla\Phi)^2\,
-{\rm e}^{-2\sigma}F^{2}
-\frac{1}{12}{\rm e}^{-4\sigma}H^2\right]}.\nonumber\\
\end{eqnarray}
Thus, one may use the conformal transformed curvature scalar
\begin{eqnarray}
{\rm e}^{2\,\sigma}{\tilde R}= { R}-2(n-1)g^{\mu\nu}
\nabla_{\nu}\nabla_{\mu}\sigma-(n-1)(n-2)g^{\mu\nu}
\nabla_{\mu}\sigma\nabla_{\nu}\sigma,
\end{eqnarray}
and chose
\begin{eqnarray}\label{conditionCT}
n\,\sigma-2\,\sigma-2\Phi=0 \rightarrow{\sigma
=\frac{2}{n-2}\Phi},\,\Phi=\frac{n-2}{2}\sigma.
\end{eqnarray}
Substituting these relations in the action above
(\ref{actioMiddle}), one arrives at
\begin{eqnarray}\label{actionEISNforString}
S=&&\int\,d^{n}x\sqrt{-g}\left[{R} -\frac{4}{n-2}(\nabla\Phi)^2
-2{\rm e}^{4\,\Phi/(n-2)}\Lambda+V(\Phi)\right.
\nonumber\\&&\left.-{\rm e}^{-4\,\Phi/(n-2)}F^{2} -\frac{1}{12}{\rm
e}^{-8\,\Phi/(n-2)}H^2 \right],
\end{eqnarray}
where it has been
dropped from this action the divergence $$ -
2(n-1)\sqrt{-g}g^{\mu\nu}\nabla_{\nu}\nabla_{\mu}\sigma=-
2(n-1)(\sqrt{-g}\sigma^{,\nu})_{,\nu},$$ and denoted $${\rm
e}^{2\sigma}U(\Phi)=V(\Phi).$$
The extremum of $S$ is achieved along the dynamical equations:
\begin{subequations}\label{EinsteinStCMgHor}
\begin{eqnarray}\label{EinsteinStCM1Hor}
&&{ G}_{\mu\nu}=-g_{\mu\nu}\,{\rm e}^{4\Phi/(n-2)}\Lambda
+\frac{1}{2}g_{\mu\nu}\,V(\Phi)+\frac{4}{n-2}
\left(\nabla_{\mu}\Phi\,\nabla_{\nu}\Phi -\frac{1}{2}
g_{\mu\nu}\nabla_{\alpha}\Phi\,\nabla^{\alpha}\Phi\right)
\nonumber\\
&& +2\,{\rm e}^{-4\Phi/(n-2)}(F_{\mu\sigma}
{F_{\nu}}^{\sigma}-\frac{1}{4} g_{\mu\nu}{F^{2}}) -\frac{1}{4}\,{\rm
e}^{-8\Phi/(n-2)}(H_{\mu\alpha\beta}
{H_{\nu}}^{\alpha\beta}-\frac{1}{6} g_{\mu\nu}H^2),
\end{eqnarray}
\begin{eqnarray}
8\,\nabla^{\nu}\nabla_{\nu}\Phi-8\,\Lambda {\rm e}^{4\Phi/(n-2)}
+4\, {\rm e}^{-4\Phi/(n-2)}\,F^{2}+\frac{2}{3}\, {\rm
e}^{-8\Phi/(n-2)}\,H^{2}+\frac{n-2}{2} \frac{d\,V}{d\,\Phi}=0,
\end{eqnarray}
\begin{eqnarray}
\nabla_{\sigma}\left({\rm
e}^{-8\Phi/(n-2)}{H}^{\alpha\beta\sigma}\right)=0,
\end{eqnarray}
\begin{eqnarray}
\nabla_{\lambda}\left({\rm e}^{-4\Phi/(n-2)} {
F}^{\lambda\epsilon}\right)+\frac{a}{12}\, {\rm
e}^{-8\Phi/(n-2)}F_{\alpha\beta}{ H}^{\alpha\beta\epsilon}=0,
\end{eqnarray}
\end{subequations}
Replacing in (\ref{EinsteinStCM1Hor}) the scalar curvature $R$,
$$R=\frac{4}{n-2}(\nabla\Phi)^2
+\frac{n-4}{n-2}\,{\rm e}^{-4\Phi/(n-2)}{ F}^2
+\frac{1}{12}\frac{n-6}{n-2}\,{\rm e}^{-8\Phi/(n-2)}{ H}^2
-\frac{n}{n-2}\,(-2\Lambda\,{\rm e}^{4\Phi/(n-2)}+V),$$ one rewrites
(\ref{EinsteinStCM1Hor}) as
\begin{eqnarray}\label{EinsteinStCMRHor}
{R}_{\mu\nu} &=&\frac{2 \Lambda}{n-2} \,g_{\mu\nu}\,{\rm
e}^{4\Phi/(n-2)}+\frac{4}{n-2} \,\nabla_{\mu}\Phi\,\nabla_{\nu}\Phi
+{\rm e}^{-4\Phi/(n-2)}\left(2\,F_{\mu\sigma}
{F_{\nu}}^{\sigma}-\frac{1}{n-2} g_{\mu\nu}\,
F^{2}\right)\nonumber\\&& +\frac{1}{4}{\rm e}^{-8\Phi/(n-2)}
\left(2\,H_{\mu\alpha\beta}{H_{\nu}}^{\alpha\beta}
-\frac{2}{3}\frac{1}{n-2} g_{\mu\nu}\,H^{2}\right) -\frac{1}{n-2}
g_{\mu\nu}\,V(\Phi).
\end{eqnarray}

\section{Dynamical equations in $2+1$ string gravity}\label{twoplusonedynString}

In the three dimensional case, the above Einstein action
(\ref{actionEISNforString}) reduces to
\begin{eqnarray}
&&S=\int{d^3x\sqrt{-g}\left[{{R}}-2{\rm e}^{4\Phi}\Lambda
-4(\nabla\Phi)^2-{\rm e}^{-4\Phi}F^{2}
-\frac{1}{12}{\rm e}^{-8\Phi}H^{2}+V(\phi)\right]},\nonumber\\
\end{eqnarray}
and the Einstein frame dynamical equations (\ref{EinsteinStCMgHor}) become
\begin{subequations}\label{EinsteinStCMgHor3}
\begin{eqnarray}\label{EinsteinStCMRHor3}
{R}_{\mu\nu} &=&{2 \Lambda} \,g_{\mu\nu}\,{\rm e}^{4\Phi}+{4}
\,\nabla_{\mu}\Phi\,\nabla_{\nu}\Phi +{\rm
e}^{-4\Phi}\left(2\,F_{\mu\sigma} {F_{\nu}}^{\sigma}- g_{\mu\nu}\,
F^{2}\right)\nonumber\\&& +\frac{1}{2}{\rm e}^{-8\Phi}
\left(\,H_{\mu\alpha\beta}{H_{\nu}}^{\alpha\beta} -\frac{1}{3}
g_{\mu\nu}\,H^{2}\right) - g_{\mu\nu}\,V(\Phi).
\end{eqnarray}
\begin{eqnarray}
8\,\nabla^{2}\Phi-8\,\Lambda {\rm e}^{4\Phi} +4\, {\rm
e}^{-4\Phi}\,F^{2}+\frac{2}{3}\, {\rm e}^{-8\Phi}\,H^{2}+\frac{1}{2}
\frac{d\,V}{d\,\Phi}=0,
\end{eqnarray}
\begin{eqnarray}
\nabla_{\sigma}\left({\rm
e}^{-8\Phi}{H}^{\alpha\beta\sigma}\right)=0,
\end{eqnarray}
\begin{eqnarray}
\nabla_{\lambda}\left({\rm e}^{-4\Phi} {
F}^{\lambda\epsilon}\right)+\frac{a}{12}\, {\rm
e}^{-8\Phi}F_{\alpha\beta}{ H}^{\alpha\beta\epsilon}=0,
\end{eqnarray}
\end{subequations}
One can recover the string dynamical
equations\label{stringdynamical} by using the conformal inverse
relations, see \cite{Eisenhart:1966book},
\begin{eqnarray}
&& {{\sigma}}_{,i;j}={{\tilde\sigma}}_{,i;j}
+2\sigma_{,i}\sigma_{,j} -{\tilde g}_{ij}({\tilde \nabla \sigma})^2,
\,{\rm e}^{-2\,\sigma}{{\tilde\sigma}^{,k}}_{\,;k}
={{\sigma}^{,k}}_{\,;k}
-(n-2)({\tilde \nabla \sigma})^2,\nonumber\\
&&
 {\stackrel{W}{{ R}_{ij}}}
 ={\tilde{\stackrel{W} {R}}}_{ij}
 +{\tilde g}_{ij} {{\tilde\sigma}^{,k}}_{;k}
 + (n-2)\left({{\tilde\sigma}}_{,i;j}+\sigma_{,i}\sigma_{,j}
 -{\tilde g}_{ij}({\tilde \nabla \sigma})^2\right),\nonumber\\
&& {\rm e}^{-2\,\sigma}{\stackrel{W}{{ R}}} = {\tilde{\stackrel{W}
{R}}}+2(n-1){{\tilde\sigma}^{,k}}_{;k} -(n-1)(n-2)({\tilde \nabla
\sigma})^2,
 \end{eqnarray}
where tilde is used to denote that covariant differentials are
constructed with $\tilde \Gamma$'s or contravariant tensor
components are build with ${\tilde g}^{\mu\nu}$. For $\sigma=2\Phi$
and $n=3$ one gets
\begin{eqnarray}\label{EinsteinStCMRn}
&& { \Phi}_{,\mu;\nu}={\tilde \Phi}_{,\mu;\nu}
+4{\Phi}_{,\mu}{\Phi}_{,\nu} -2{\tilde g}_{\mu\nu}({\tilde \nabla}\Phi)^2,\nonumber\\
&& {{ \Phi}^{;\alpha}}_{;\alpha}={\rm e}^{4\Phi}\, \left({{\tilde
\Phi}^{\,;\alpha}}_{;\alpha} -2({\tilde \nabla}\Phi)^2\right),
\nonumber\\
&&{R}_{\mu\nu}={\tilde R}_{\mu\nu} +2{\tilde \Phi}_{,\mu;\nu}
+2{\tilde g}_{\mu\nu}{{\tilde \Phi}^{;\alpha}}_{;\alpha}
+4{\Phi}_{,\mu}{\Phi}_{,\nu}-
4{\tilde g}_{\mu\nu}({\tilde \nabla}\Phi)^2,\nonumber\\
&& R={\rm e}^{4\Phi}\left({\tilde R} +8\,{{\tilde
\Phi}^{;\alpha}}_{\,;\alpha}-8\,({\tilde \nabla}\Phi)^2\right),
\end{eqnarray}
and conformally transforming the above dynamical equations
(\ref{EinsteinStCMgHor3}), using relations (\ref{EinsteinStCMRn}), one
gets the barred dynamical equations of the $2+1$ string theory under
consideration
\begin{subequations}\label{FieldEQstring}
\begin{eqnarray}
&&S=\int{d^3x\sqrt{-{\tilde g}}{\rm e}^{-2\Phi}\left[{\tilde {R}}-2\Lambda
+4({\tilde \nabla}\Phi)^2-{\tilde F}^{2}-\frac{1}{12}{\tilde H}^{2}+U(\phi)\right]},\nonumber\\
\end{eqnarray}
\begin{eqnarray}\label{tildeEinsHor}
&&{\tilde R}_{\mu\nu}+2\nabla_{\mu}\nabla_{\nu}\Phi
-2F_{\mu\alpha}F_{\nu\beta}{\tilde g}^{\alpha\beta}- \frac{1}{4}{
H}_{\mu\alpha\beta}{ H_{\nu\gamma\lambda}}{\tilde
g}^{\gamma\alpha}{\tilde g}^{\lambda\beta}\nonumber\\&& +{\tilde g}_{\mu\nu}\left(-2\Lambda
+U+{\tilde F}^{2}+\frac{{\tilde H}^{2}}{6}
+2{\tilde \nabla}^2\Phi-4({\tilde \nabla}\Phi)^2\right)=0,
\end{eqnarray}
\begin{eqnarray}\label{tildeEinsHorScal}
{2}{\tilde\nabla}^2\Phi
-4 ({\tilde\nabla}\Phi)^2+\,{
\tilde F}^{2}-2\Lambda+\frac{1}{6}{\tilde H}^2+U(\Phi)
+\frac{1}{4}\frac{d\,U}{d\,\Phi} =0,
\end{eqnarray}
\begin{eqnarray}\label{BfieldHOR}
{\tilde\nabla}_{\sigma}({\rm e}^{-2\Phi}{\tilde
H}^{\mu\nu\sigma})=0,
\end{eqnarray}
\begin{eqnarray}\label{AfieldHOR}
{\tilde\nabla}_{\lambda}\left({\rm e}^{-2\Phi}{\tilde
F}^{\lambda\epsilon}\right) +a\frac{1}{12}{\rm e}^{-2\Phi}{\tilde
F}_{\nu\lambda} \,{\tilde H}^{\epsilon\nu\lambda}=0.
\end{eqnarray}
\end{subequations}

\section{Horne-Horowitz black string}\label{Horne-HorowitzSTR}

In 1991, Horne and Horowitz published an exact string black hole
solution in three dimensions ~\cite{HorneHorowitzNPB91} for the full
string theory of Sec.~\ref{StringEqHOR}, endowed with mass, axion
charge per unit length, the asymptotic value of the dilaton, and a
cosmological constant. This string solution is given by
\begin{eqnarray}\label{HorHor}
&&{\stackrel{S}{g}}=-(1-\frac{M}{r})dt^2 +(1-\frac{Q^2}{M\,r})dx^2
+(1-\frac{M}{r})^{-1}(1-\frac{Q^2}{M\,r})^{-1}\frac{k\,dr^2}{8r^2},
\nonumber\\&&
H_{rtx}=\frac{Q}{r^2},\,\,\phi=\ln{r}+\frac{1}{2}\ln{\frac{k}{2}}.
\end{eqnarray}
The identification of the functions appearing in the action
(\ref{actionHor}) and equations (\ref{einsteinHORdyn}) with the ones
of \cite{HorneHorowitzNPB91}
corresponds to  $\phi=-2\Phi,$ and
$\frac{8}{k}=-2\Lambda=\frac{4}{l^2},\, k=2\,l^2$,
where $l$ has dimension of length.\\

Accomplishing in the above--mentioned metric a conformal
transformation ${\stackrel{E}{g}}_{\mu\nu}={\rm e}^{-2\sigma}
{\stackrel{S}{g}}_{\mu\nu}=\frac{k\,r^2}{2}{\stackrel{S}{g}}_{\mu\nu},$
where superscripts $E$ and $S$ stand for Einstein and String
respectively, one arrives at the corresponding solution in the
Einstein frame, namely
\begin{eqnarray}\label{HorHorEF}
&&{\stackrel{E}{g}} =-\frac{k\,r^2}{2}(1-\frac{M}{r})dt^2
+\frac{k\,r^2}{2}(1-\frac{Q^2}{M\,r})dx^2+
(1-\frac{M}{r})^{-1}(1-\frac{Q^2}{M\,r})^{-1}\frac{k^2\,dr^2}{16},
\nonumber\\&& H_{rtx}=\frac{Q}{r^2},\,\Phi
=-\frac{1}{2}\ln{r}-\frac{1}{4}\ln{\frac{k}{2}},
\end{eqnarray}
fulfilling the dynamical equations (\ref{EinsteinStCMgHor}) for
dimension $n=3$.

By accomplishing a $SL(2,R)$ transformation of the
Killing coordinates
\begin{eqnarray}\label{GENmetricHH}
&&t=\alpha\,T+\beta\,\phi,\,\,\Delta:=\alpha\delta-\beta\gamma.
\nonumber\\&& x=\gamma\,T+\delta\,\phi,
\end{eqnarray}
one arrives at a stationary HH--string
solution\label{stationaryHH--string}
\begin{eqnarray}\label{HorHorEF}
&&{\stackrel{E}{g}}=-\frac{k\,r^2}{2} \left[(1-\frac{M}{r})\alpha^2
-(1-\frac{Q^2}{M\,r})\gamma^2\right]dT^2
+{k\,r^2}\left[(1-\frac{Q^2}{M\,r})
\gamma\delta-(1-\frac{M}{r})\alpha\beta\right]dT\,d\phi\nonumber\\&&
+\frac{k\,r^2}{2}\left[(1-\frac{Q^2}{M\,r})\delta^2
-(1-\frac{M}{r})\beta^2\right]d\phi^2+
(1-\frac{M}{r})^{-1}(1-\frac{Q^2}{M\,r})^{-1}\frac{k^2\,dr^2}{16},
\nonumber\\&& H_{rtx}=\frac{Q}{r^2},\,\Phi
=-\frac{1}{2}\ln{r}-\frac{1}{4}\ln{\frac{k}{2}},\frac{8}{k}=-2\Lambda==\frac{4}{l^2}.
\end{eqnarray}
In the literature one frequently encounters the $SL(2,R)$
transformation
\begin{eqnarray}
t={\frac {T}{\sqrt {1-{\frac {{\omega}^{2}}{{l}^{2}}}}}}-\omega\,
{\frac {\phi}{\sqrt {1-{\frac {{\omega}^{2}}{{l}^{2}}}}}},\, x=-
\frac{\omega}{{l}^ {2}}\,{\frac {T}{\sqrt {1-{\frac
{{\omega}^{2}}{{l}^{2}}}}}}+ {\frac {\phi}{\sqrt {1-{\frac
{{\omega}^{2}}{{l}^{2}}}}}},
\end{eqnarray}
which yields
\begin{eqnarray}
&&g=-{\frac { \left( r-M \right)  \left( rM-{Q}^{2} \right) rk
\left( {l}^{2}-{\omega}^{2} \right) }{2\,{l}^{2} \left(
-{\omega}^{2}rM+{\omega} ^{2}{M}^{2}+rM-{Q}^{2} \right) }}dT^2+
(1-\frac{M}{r})^{-1}(1-\frac{Q^2}{M\,r})^{-1}\frac{k^2\,dr^2}{16}
\nonumber\\&& +{\frac {{l}^{2}kr \left(
-{\omega}^{2}rM+{\omega}^{2}{M}^{2}+rM-{ Q}^{2} \right) }{2M \left(
{l}^{2}-{\omega}^{2} \right) }} \left(d\phi-{\frac {\omega\, \left(
-{l}^{2}Mr+{l}^{2}{M}^{2}+rM-{Q}^{2} \right) }{{l}^{2} \left(
-{\omega}^{2}rM+{\omega}^{2}{M}^{2}+rM-{Q}^{2}
 \right) }}
dT\right)^2, \nonumber\\&&
H_{rtx}=\frac{Q}{r^2},\,\Phi=-\frac{1}{2}\ln{r}
-\frac{1}{4}\ln{\frac{k}{2}}.
\end{eqnarray}
The evaluation of the quasilocal mass, energy and momentum is done
using the Brown--York approach~\cite{BrownY93}, this yields
\begin{subequations}
\begin{eqnarray}
J(r\rightarrow \infty)=2\omega\,{l}^{2} \,{\frac { \left( {M}^{2}
-{Q}^{2}\right) }{ \left( { l}^{2}-{\omega}^{2} \right) M}},
\end{eqnarray}
\begin{eqnarray}
\epsilon(r\rightarrow \infty)
=-1/2\,{\frac {1}{r\pi\,{l}^{2}}}-1/4\,{\frac { \left( M-\omega\,Q
 \right)  \left( M+\omega\,Q \right) }{M{l}^{2}\pi\, \left( \omega-1
 \right)  \left( \omega+1 \right) {r}^{2}}}-\epsilon_0,
\end{eqnarray}
\begin{eqnarray}E(r\rightarrow \infty)={\frac {2\,\sqrt {1-{\omega}^{2}}}{\sqrt {{l}^{2}
-{\omega}^{2}}}}-2\pi\,K(r)\epsilon_0,
\end{eqnarray}
\begin{eqnarray}
M_{BY}(r\rightarrow \infty)=4r+4\,l\,(-2r^2+M_0l^2),
\end{eqnarray}
\begin{eqnarray}\epsilon_0=-{\frac {1}{\pi\,l}}
+\frac{1}{2}\,{\frac {{\it M_0\l}}{\pi\,{r}^{2} }},
\end{eqnarray}
\end{subequations}
$\omega$ is interpreted as a rotating parameter, the mass function
increases as the radial coordinate approach spatial infinity.
Moreover various pathologies take place at this location.

\section{Horowitz--Welch black string}\label{Horne-HorowitzSTR}

In 1993, Horowitz and Welch published an exact string black hole
solution in three dimensions ~\cite{Horowitz:1993jc} for the low
energy string theory
\begin{eqnarray}\label{actionHorWel}
S=\int{d^{\,3}x\sqrt{- g}{\rm e}^{-2\Phi}\left[{ R} -2 \Lambda+4
(\nabla\Phi)^2 -\frac{1}{12}H^{2}\right]},\,\Lambda=-2/ k_{HW},
\end{eqnarray}
of Sec.~\ref{StringEqHOR}, endowed with mass, angular momentum,
axion charge per unit length, and a negative cosmological constant.
This string solution is given by a modified BTZ black hole to a
$2+1$ string theory with vanishing scalar $\Phi$ and electromagnetic
$F_{\alpha\beta}$ fields; in this last case the field equations
(\ref{FieldEQstring}) become
\begin{subequations}\label{FieldEQstringW}
\begin{eqnarray}\label{tildeEinsHorW}
{ R}_{\mu\nu}- \frac{1}{4}{ H}_{\mu\alpha\beta}{
H_\nu}^{\alpha\beta}=0,
\end{eqnarray}
\begin{eqnarray}\label{tildeEinsHorScalW}
-2\Lambda+\frac{1}{6}H^2 =0,
\end{eqnarray}
\begin{eqnarray}\label{BfieldHORW}
\nabla_{\sigma}(H^{\mu\nu\sigma})=0.
\end{eqnarray}
\end{subequations}
The totally anti--symmetric tensor $H_{\mu\nu\alpha}$ has to be
proportional to the volume three form $\epsilon_{\mu\nu\alpha}$,
because of the equation (\ref{BfieldHORW}), the proportionality
factor ought to be a constant, hence one may chose
\begin{eqnarray}\label{BfieldHORW1}
H_{\mu\nu\sigma}=\alpha\frac{1}{l}\epsilon_{\mu\nu\sigma}.
\end{eqnarray}
Taking into account the properties of $\epsilon$
\begin{eqnarray}\label{BfieldHORW2}
\epsilon_{\alpha\nu\sigma}\epsilon^{\beta\nu\sigma}=-2{\delta^\beta}_{\alpha},
\end{eqnarray}
therefore
\begin{eqnarray}\label{BfieldHORW2}
H_{\alpha\nu\sigma}H^{\beta\nu\sigma}
=-2\frac{\alpha^2}{l^2}{\delta^\beta}_{\alpha},\,H^2=-6\frac{\alpha^2}{l^2}.
\end{eqnarray}
Consequently, from (\ref{tildeEinsHorW}) one has
\begin{eqnarray}\label{tildeEinsHorW}
{ R}_{\mu\nu}- \frac{1}{4}{ H}_{\mu\alpha\beta}{
H_\nu}^{\alpha\beta}=0\rightarrow{{ R}_{\mu\nu}=-\frac{2}{l^2}{
g}_{\mu\nu},\,\alpha=2}
\end{eqnarray}
\begin{eqnarray}\label{tildeEinsHorScalW}
-2\Lambda+\frac{1}{6}H^2
=0\rightarrow{\Lambda=-\frac{\alpha^2}{2l^2}=-\frac{2}{l^2},\, k_{HW}=l^2}
\end{eqnarray}
On the other hand, $\bm H=\bm{dB}$, thus
\begin{eqnarray}\label{tildeEinsHorScalWx}
H_{\alpha\beta\gamma}=3B_{[\alpha\beta,\gamma]}
=\frac{2}{l}\epsilon_{\alpha\beta\gamma}.
\end{eqnarray}
As concluded in~\cite{Horowitz:1993jc}: ``Thus every solution to
three dimensional general relativity with negative cosmological
constant is a solution to low energy string theory with: $\Phi=0$,
$H_{\alpha\beta\gamma}=\frac{2}{l}\epsilon_{\alpha\beta\gamma}$, and
$\Lambda=-\frac{2}{l^2}$.''

In particular in~\cite{Horowitz:1993jc} is established that the
BTZ black hole metric
\begin{eqnarray}\label{HorWELCHmetBTZ}
&&{g}=-(\frac{r^2}{l^2}-{M})dt^2-J\,dt\,d\phi +{r}^2d\phi^2
+(\frac{r^2}{l^2}-{M}+\frac{J^2}{4l^2})^{-1}dr^2,
\end{eqnarray}
in the presence of an anti--symmetric $B$ field
\begin{eqnarray}\label{HorWELCHmetB}
B_{\phi\,t}=\frac{r^2}{l},\,\bm H=\bm{dB},
\end{eqnarray}
is a solution of the string theory with a zero scalar field
$\Phi$.\\
By a target space duality transformation (13)
of~\cite{Horowitz:1993jc}, which referred to \cite{Buscher88}, which means
that from a given solution $(g_{\mu\nu},\,B_{\mu\nu},\,\Phi)$ independent on one variable, say $x$, one generates
a new  solution $(\tilde{g}_{\mu\nu},\,\tilde{B}_{\mu\nu},\,\tilde{\Phi})$ with
\begin{eqnarray}\label{HorWELCHmetBT}
&&\tilde{g}_{xx}=\frac{1}{{g}_{xx}},\,\tilde{g}_{x\alpha}
=\frac{B_{x\alpha}}{{g}_{xx}},\nonumber\\&&
\tilde{g}_{\alpha\beta}={g}_{\alpha\beta}
-\frac{1}{{g}_{xx}}({g}_{x\alpha}\,{g}_{x\beta}
-{B}_{x\alpha}\,{B}_{x\beta}),\nonumber\\&&
\tilde{B}_{x\alpha}=\frac{{g}_{x\alpha}}{{g}_{xx}},\,
\tilde{B}_{\alpha\beta}={B}_{\alpha\beta}
-2\frac{{g}_{x[\alpha}{B}_{\beta]\,x}}{{g}_{xx}},\nonumber\\&&
\tilde{\Phi}={\Phi}-\frac{1}{2}\ln{{g}_{xx}},
\end{eqnarray}
where $\alpha$  and $\beta$ run over all directions except $x$.
Applying this transformation  to expressions
(\ref{HorWELCHmetBTZ}) and (\ref{HorWELCHmetB}), along the  coordinate symmetry $\phi$, one gets (14)
of~\cite{Horowitz:1993jc}, namely
\begin{eqnarray}\label{HorWELCHmet}
&&{\stackrel{S}{g}}=(M-\frac{J^2}{4r^2})dt^2
+\frac{2}{l}dt\,d\phi+\frac{1}{r^2}\,d\phi^2
+(\frac{r^2}{l^2}-M+\frac{J^2}{4r^2})^{-1}dr^2,
\end{eqnarray}
which, once diagonalized by means of a $SL(2,R)$ coordinate
transformation
\begin{eqnarray}\label{HorWELCHtrnscoor}
&&t=-\frac{l}{\sqrt{r_{+}^2-r_{-}^2}}\tilde{t}
+\frac{l}{\sqrt{r_{+}^2-r_{-}^2}}\tilde{x},\nonumber\\&&
\phi=\frac{r_{+}^2}{\sqrt{r_{+}^2-r_{-}^2}}\tilde{t}
-\frac{r_{-}^2}{\sqrt{r_{+}^2-r_{-}^2}}\tilde{x},
\end{eqnarray}
and the $r$--coordinate transformation
\begin{eqnarray}\label{HorWELCHtrnscoorRad}
&&r^2=l\tilde{r}
\end{eqnarray}
yields the string solution derived in~\cite{HorneHorowitzNPB91}, see
also Sec. \ref{Horne-HorowitzSTR}. Dropping primes, it becomes
\begin{eqnarray}\label{HorHor}
&&{\stackrel{S}{g}}=-(1-\frac{\mathcal{M}}{r})dt^2
+(1-\frac{\mathcal{Q}^2}{\mathcal{M}\,r})dx^2
+(1-\frac{\mathcal{M}}{r})^{-1}(1-\frac{\mathcal{Q}^2}
{\mathcal{M}\,r})^{-1}\frac{l^2\,dr^2}{4r^2},\, \nonumber\\&&
B_{xt}=\frac{\mathcal{Q}}{r},\,\,\phi=-\frac{1}{2}\ln{(r\,l)},
\end{eqnarray}
where $\mathcal{M}={r_{+}^2}/l$ and $\mathcal{Q}=J/2$.\\
The
identification of the functions appearing in the action
(\ref{actionHor}) and equations (\ref{einsteinHORdyn}) with the ones
of \cite{HorneHorowitzNPB91},
\begin{eqnarray}\label{actionHorHorcom}
S=\int{d^{\,3}x\sqrt{- g}{\rm e}^{\phi}\left[{ R} -2 \Lambda+
(\nabla\Phi)^2 -\frac{1}{12}H^{2}\right]},\, \Lambda=-4/k_{HH},
\end{eqnarray}
requires that $\phi=-2\Phi_{HW},$ and $k_{HH}=2\,k_{HW}$, $k_{HW}=l^2$, $\Lambda=-2/l^2$,
where the subscripts are in correspondence
with the initial of the author's family name.

\section{Chan--Mann string solution}\label{Chan--Mannstring}

Chan and Mann~\cite{Chan:1994qa}, see also \cite {Chan:1995wj},
derived a class of solutions to dilaton minimally
coupled to $2+1$ Einstein--Maxwell gravity. There is a subclass
of solutions allowing an interpretation from the viewpoint of the low energy
$2+1$ string theory for specific values of the charged dilaton solution.
First, we derive the dilaton solution. Next, assigning specific values to
constant characterizing the charged dilaton, the correspondence with
the string theory developed in the previous section is established.

\subsection{Einstein--Maxwell--scalar field equations}\label{sec:cyclic}
 The action considered in~\cite{Chan:1994qa}, CM~(1), for a
 (2+1)-dimensional gravity is given by
 \begin{equation}\label{action}
 \mathcal{S}=\int{d^3x}\sqrt{-g}\left[
 R-\frac{B}{2}\nabla_{\mu}{\Psi}\,\nabla^{\mu}{\Psi}
 +2\,{\rm e}^{b\Psi}\Lambda_{CM}-{\rm e}^{-4\,a\,\Psi}\,{ F^2}\right],
 \end{equation}
 where $\Lambda_{CM}, b$ are arbitrary at this stage parameters,
 ${\Psi}$ is the massless
 minimally coupled scalar field, $R$ is the scalar curvature,
 and  $F^2=F_{\mu\,\nu}\,F^{\mu\,\nu}$ the electromagnetic invariant.
 The variations of this action yield the dynamical equations
  \begin{eqnarray}\label{Einsteinfield}
 &&{R_{\mu\nu}}= \frac{B}{2}
 \nabla_{\mu}{\Psi}\,\nabla_{\nu}{\Psi}-2{g_{\mu\nu}}{\rm e}^{\,b\,\Psi}\Lambda_{CM}
 +2\,{\rm e}^{-4\,a\,\Psi}\left({F_\mu}^\alpha\,{F_\nu}_\alpha-g_{\mu\,\nu}\,F^2\right),
 \nonumber\\
 &&
 \frac{B}{2}{\nabla}^{\mu}{\nabla}_{\mu}{{\Psi}}
 +b\,{\rm e}^{\,b\,\Psi}\Lambda_{CM}+2\,a\,{\rm e}^{-4\,a\,\Psi}\,{ F^2}
 =0, \nonumber\\
 &&
 \nabla^{\mu}\left({\rm e}^{-4\,a\,\Psi}\,F_{\mu\,\nu}\right)=0.
 \end{eqnarray}

 \subsection{ General static cyclic symmetric black hole solution
 coupled to a scalar field $\Psi(r)=k\,{\rm \ln}(r)$}\label{ScalarStatic}

 The static cyclic symmetric metric in the $2+1$
 Schwarzschild coordinate frame is given by
 \begin{equation}\label{metricscLNK_rW_0}
 \bm{g}=-N (r)^2\bm{dt}^2+\frac{\bm{dr}^2}{L(r)^2}
 +r^2\bm{d\phi}^2.
 \end{equation}
 The electromagnetic field equations for the tensor
 field $F_{\mu\nu}=2F_{tr}\delta^t_{[\mu}\delta^r_{\nu]}$,
 and the dilaton $\Phi(r)=k\,{\rm \ln}(r)$ becomes
 \begin{eqnarray}\label{scalarfeqr}
 EQ_F=\frac{d}{dr}\frac {F_{tr} \, L \, r^{-4\,a\,k+1}}{N}
  \rightarrow F_{tr} = {\it Q}\frac{N}{L}{r}^{4\,ak-1}.
 \end{eqnarray}
 The simplest Einstein equations occurs to
 be ${\it R_{tt}}+{\it R_{rr}}\,L^2\,N^2$, which yields
 \begin{eqnarray}
{\frac {1}{N }}{\frac {d}{dr}}N-{
\frac {1}{L }}{\frac {d}{dr}}L -\frac{1}{2}
\,{\frac {B{k}^{2}}{{r}}}=0,
 \end{eqnarray}
 thus one gets
 \begin{eqnarray}
 N(r)={\it C_N}\,L \left( r \right) {r}^{B\,{k}^{2}/2}.
 \end{eqnarray}
 On the other hand, the equation ${\it R_{\phi\phi}}$
 gives a first order equation for $Y \left( r \right)=L^2$, namely
 \begin{eqnarray}\label{eqL2sol}
 &&
 {\frac {d}{dr}}Y \left( r \right) +\frac{1}{2}\,{\frac {B{k}^{2}Y \left( r
 \right) }{r}}+2\,{\frac { {r}^{4\,ak} {{\it Q}}^{2}}
 {r}}-2\,\Lambda_{CM}\,{r}^{bk+1}=0
 \end{eqnarray}
 integrating one obtains
 \begin{eqnarray}\label{L2sol}
 L(r)^2=Y \left( r \right) =-4\,{\frac {{r}^{4\,ak}{{\it Q}}^{2}}{B{k}^{
2}+8\,ak}}+4\,{\frac {\Lambda_{CM}\,{r}^{2+bk}}{4+\,B{k}^{2}+2bk}}+{r}^{-
1/2\,B{k}^{2}}{\it C_1}
\end{eqnarray}

 Substituting this expression of $Y \left( r \right)$ into the
 remaining scalar field equation
\begin{eqnarray}
 {\frac {d}{dr}}Y \left( r \right) +\frac{1}{2}\,{\frac {B{k}^{2}Y \left( r
 \right) }{r}}-8\,{\frac {
{r}^{4\,ak}\,a{{\it Q}}^{2}}{B\,k\,r}}+2\,{\frac {b\Lambda_{CM}\,{r}^{bk+1}}{Bk}}=0,
 \end{eqnarray}
 one arrives at
 relationships between constants, namely
 \begin{eqnarray}
 a=-\frac{1}{4\,B\,k},\,b=-B\,k.
 \end{eqnarray}
 \noindent\\
 Therefore, the general charged dilaton  static solution can be given as
 \begin{eqnarray}\label{staticGADxe}
 &&
 \b{g}=-{\it C_N}^{2}\,{r}^{B\,{k}^{2}}
 \,L(r)^2{dt}^2+\frac{{dr}^2}{L(r)^2}+r^2{d\phi}^2,
 \nonumber\\
 &&L(r)^2=\left({r}^{B\,{k}^{2}/2}{\it C_1}
 +4\,{\frac {{r}^{2}\Lambda_{CM}}{4-B\,{k}^{2}}}+4\,{\frac {{{\it Q}}^{2}}{B{
k}^{2}}}
 \right) {r}^{-B\,{k}^{2}},\,{k}^{2}\neq\frac{4}{B},
 \nonumber\\
 &&
 F_{\mu\nu}=2F_{tr}\delta^t_{[\mu}\delta^r_{\nu]},
 \,F_{tr}={\it Q}{\it C_N}\,{r}^{-1/2\,B{k}^{2}-1}=-A_{t,r}
 \rightarrow\,A_{t}
 =2\frac{{\it Q}{\it C_N}}{B\,k^2} \,{r}^{-B\,{k}^{2}/2},\nonumber\\
 &&
 \Psi(r)=k\,{\rm \ln}{(r)},
 \end{eqnarray}
 endowed with four relevant parameters: in particular, one
 may identify the mass $M=-{\it C_1}$,
 cosmological constant $\Lambda_{CM}\rightarrow\pm\frac{1}{l^2}$,
 dilaton parameter  $k$, and the charge $
 {\it Q}$. The constant ${\it C_N}$ can be absorbed by scaling
 the coordinate $t$, thus it can be equated to unit,
 ${\it C_N}\rightarrow 1$. Moreover, one has to set
 the charge $Q$ to zero, $Q=0$, when looking for the
 limiting solutions for vanishing dilaton $k=0$, which are
 just the dS and AdS soutions with parameters ${\it C_1}=\pm M$
 respectively, and ${\it C_N}=1$. There is no static
 electrically charged limit of this solution for vanishing dilaton field.

 The constant $\Lambda_{CM} $ can be
 equated to minus the standard cosmological
 constant $\Lambda_{s}=\pm \frac{1}{l^2}$;
 indeed, by setting in (\ref{staticGADxe})
\begin{eqnarray}
 &&\Lambda_{CM}=\pm \frac{1}{l^2}\alpha^2,
 \,r\rightarrow r\alpha^{2/(B\,k^2)},\,\phi
 \rightarrow \phi\,\alpha^{-2/(B\,k^2)},
 {\it Q}\rightarrow {\it Q}\,\alpha^{(1+2/(B\,k^2))},
 \nonumber\\
 &&
 {\it C_1}\rightarrow {\it C_1}\,\alpha^{1+4/(B\,k^2)},
 {\it C_N}\rightarrow {\it C_N}\,\alpha^{-(1+2/(B\,k^2))},
\end{eqnarray}
one arrives at the metric (\ref{staticGADxe}) with $\Lambda_{CM}=\pm\frac{1}{l^2}$.
Notice that the $\Lambda_{CM}$ used by in Chan--Mann works,
when considered as a cosmological
constant, differs from the standard
cosmological constant $\Lambda_{s}=\pm \frac{1}{l^2}=-\Lambda_{CM}$,
 where $+$ and $-$ stand correspondingly for de Sitter and
 Anti de Sitter (AdS) in $\Lambda_{s}$.\noindent\\

The string solution derived in~\cite{Chan:1994qa}, for $B=8,k=-1/2,
a=1,b=4$, which fulfills the Einstein string equations
(\ref{EinsteinStCMgHor3}) for
$\Lambda =\Lambda_{s}=\pm\frac{1}{l^2}=-\Lambda_{CM}$ , in the case of
vanishing $H$, can be given as
\begin{eqnarray}\label{staticGADx}
 &&
 {g}_{E}=-{r}^{{2}}
 \,L(r)^2{dt}^2+\frac{{dr}^2}{L(r)^2}+r^2{d\phi}^2,
 \nonumber\\
 &&L(r)^2=\left({r}\,{\it C_1}
 -2\,{r}^{2}\Lambda_{s}+2\,{\it Q}^{2}
 \right) {r}^{-{2}},
 \nonumber\\
 &&
 F_{\mu\nu}=2F_{tr}\delta^t_{[\mu}\delta^r_{\nu]},
 \,F_{tr}=\frac{{\it Q}}{r^2}=-A_{t,r}
 \rightarrow\,A_{t}=\frac{{\it Q}}{r},\nonumber\\
 &&
 \Psi(r)=-1/2\,{\rm \ln}{(r)}.
 \end{eqnarray}
Under the conformal transformation
\begin{eqnarray} {\tilde g}_{\mu\nu}
={\rm e}^{4\,\Psi(r)}{g}_{\mu\nu} =\frac{1}{r^2}{g}_{\mu\nu}
\end{eqnarray}
it becomes
\begin{eqnarray}\label{staticGADxs}
 &&
 {g}_{S}=-\left({r}\,{\it C_1}
 -2\,{r}^{2}\Lambda_{s}+2\,{\it Q}^{2}
 \right){r}^{{-2}}
 \,{{dt}^2}+\frac{{dr}^2}{\left({r}\,{\it C_1}
 -2\,{r}^{2}\Lambda_{s}+2\,{\it Q}^{2}
 \right) }+{d\phi}^2,
 \nonumber\\
 &&
 F_{\mu\nu}=2F_{tr}\delta^t_{[\mu}\delta^r_{\nu]},
 \,F_{tr}=\frac{{\it Q}}{r^2}=-A_{t,r}
 \rightarrow\,A_{t}=\frac{{\it Q}}{r},\nonumber\\
 &&
 \Psi(r)=-1/2\,{\rm \ln}{(r)},
 \end{eqnarray}
this is a solution of the equations
(\ref{FieldEQstring}) of the $2+1$ string theory.

Moreover, subjecting the metric (\ref{staticGADxs}) to $SL(2,R)$
transformations of the Killing coordinates
\begin{eqnarray}
t=\alpha\,\tau+\beta\,\theta,\,\, \phi=\gamma\,\tau+\delta\,\theta
\end{eqnarray}
one arrives at a rotating charged string
solution\label{CMrotatingstring}, namely
\begin{eqnarray}\label{staticGADx}
{g_{S}} &&=-\left(\alpha^2\, \mathcal{L}^2/{r}^{2}-\gamma^2\right)
 {{d\tau}^2}
 -2\left(\alpha\beta\,\mathcal{L}^2/{r}^{2}
 -\gamma\,\delta\right) {d\theta}\, {d\tau}
+\left(\delta^2-\beta^2\,\,\mathcal{L}^2/{r}^{2}\right)
  {d\theta}^2+\frac{ {dr}^2}{\mathcal{L}^2 },
 \nonumber\\
 &&
 F_{\mu\nu}=2\alpha\,F_{tr}\delta^\tau_{[\mu}\delta^r_{\nu]}
 -2\beta\,F_{tr}\delta^\theta_{[\mu}\delta^r_{\nu]},
 \,F_{tr}:=\frac{{\it Q}}{r^2}=-A_{t,r}
 \rightarrow\,A_{t}=\frac{{\it Q}}{r},\nonumber\\
 &&
 \Psi(r)=-1/2\,{\rm \ln}{(r)},\,\,
 \mathcal{L}^2:={r}\,{\it C_1}
 -2\,{r}^{2}\Lambda_{s}+2\,{\it Q}^{2}.
 \end{eqnarray}
In particular when $\Lambda_{CM}=1/l^2$, AdS branch, a usual choice of
the $SL(2,R)$ transformations is given by
 \begin{eqnarray}
t=\frac{\tau}{\sqrt{1-\omega^2/l^2}}
-\omega\frac{\theta}{\sqrt{1-\omega^2/l^2}}, \,\,
\phi=-\frac{\omega}{l^2}\frac{\tau}{\sqrt{1
-\omega^2/l^2}}+\frac{\theta}{\sqrt{1-\omega^2/l^2}},
\end{eqnarray}
where $\omega$ stands for the rotation parameter. This metric can be
written as
\begin{eqnarray}
{g_{S}} &&=-\frac{\mathcal{L}^2}{r^2}
\frac{1-{\omega^2}/{l^2}}{1-\omega^2\, \mathcal{L}^2/r^2}
 { {d\tau}^2}
 +\frac{1-\omega^2\,\mathcal{L}^2/r^2}{1-{\omega^2}/
 {l^2}}\left( {d\theta}-\frac{\omega}
 {l^2}\frac{1-{\omega^2}/{l^2}}
 {1-\omega^2\,\mathcal{L}^2/r^2}\,
 {d\tau}\right)^2
 +\frac{ {dr}^2}{\mathcal{L}^2},\nonumber\\
\end{eqnarray}
and is endowed with four parameters: the mass $M=-{\it C_1}$, charge
${\it Q}$, rotation $\omega$, and cosmological constant
$\Lambda_{s}=\pm1/l^2.$

It is worthy to point out that string solutions
one can find in various dimensions, for instance, the works \cite{WittenPRD91}
\cite{MandalSW_MPLA91}, and \cite{MakiSH_CQG93}, among others.

\section{Acknowledgments}
This work has been partially carried out during a sabbatical year at the
Physics Department--UAMI, and has been supported by Grant
CONACyT 178346.

\end{document}